\documentclass[11pt,a4paper]{article}

\usepackage[super]{natbib}

\begin{titlepage}
\title{ Lorentz Transformation Equations in Galilean Form}
\author{Sadanand D. Agashe\\Department of Electrical Engineering\\Indian Institute of Technology Bombay, Powai, \\Mumbai-76 \\India - 400076\\email: eesdaia@ee.iitb.ac.in}
\date{}
\end{titlepage}

\begin{document}
\maketitle

\begin{abstract}
 Using the definition of ``position'' given in an earlier paper, we show that the Lorentz transformation equations for position can be put in a particularly simple form which could be said to be ``Galilean''. We emphasize that two different reference frames use their individual definition of position and distance. This fact gets obscured in the usual ``rectangular Cartesian co-ordinate system'' approach.
\end{abstract}

\section{WAS  EINSTEIN'S KINEMATICS INCOMPLETE?}

Einstein, in his pioneering paper \cite{EinsteinA1905}, insisted that when talking about motion, we must give a \emph{physical} meaning to ``time''. He showed how this can be done by introducing his idea of ``synchronized clocks''. However, he took for granted the concept of ``position''. He wrote:
\begin{quote} Let us take a system of co-ordinates in which the equations of Newtonian mechanics hold good. \ldots If a material point is at rest relatively to this system of co-ordinates, its position can be defined relatively thereto by the employment of rigid standards of measurement and the methods of Euclidean geometry, and can be expressed in Cartesian co-ordinates.
\end{quote}
Thus, he did not describe how position could be defined in a general, physically meaningful way.  Nobody subsequently has done so. A little further in \cite{EinsteinA1905}, he said:
\begin{quote}
Let us in ``stationary'' space take two systems of co-ordinates, i.e.,two systems, each of three rigid material lines, perpendicular to one another and issuing from a point.
\end{quote}
Are the co-ordinate axes, then, material bodies?  Do we need to think of what might happen to them when they move? Or are they merely conceptual? In a sense, then, Einstein's Kinematics was incomplete.

\section{EINSTEIN'S KINEMATICS COMPLETED}

In a recent paper \cite{Agashe1}, we suggested how Einstein's Kinematics could be completed.  We proposed that an ``observation system'' could have the following ingredients.  An observer, $S$, say, capable of sending and receiving light signals, is equipped with a \emph{single} clock and \emph{three} passive ``reflecting'' stations, say, $S_1,S_2,S_3$. Using a radar-like approach, the observer could obtain data sets as follows.  He sends a signal in all directions at a time $t_0$ in his clock, and then records the times of arrivals of the echoes of this signal by reflection in the following four different ways: time $t_0'$ of arrival after reflection at the place P, say, of an event being observed (path $SPS$); time $t_1$ of arrival after reflection at the event P first, followed by a reflection at the station $S1$ back to $S$ (path $SPS_1S$); similarly, time instants $t_2$,$t_3$.  Thus, he would obtain 5-tuples of data items of time, ${t_0,t_0',t_1,t_2,t_3}$.  From each 5-tuple, he is to decide \emph{by definition} what could be meaningfully called the ``time of occurrence'' and ``place'' of the event.   Following Einstein, we chose $t=\frac{t_0+t_0'}{2}$ as the \emph{definition} of the time of occurrence.

There remained the problem of deciding what could meaningfully be called the place of the event.  Here, we proposed to go beyond the classical 3-dimensional rectangular Cartesian co-ordinate system idea which was only conceptual; we chose instead to think of the place of an event as an element of a 3-dimensional vector space to be equipped with a suitable scalar or inner product.  The place of an event could thus be thought of as a ``position vector''.  Any 3-dimensional vector space, $V$, say, would do.  (Today, we know the possible advantages of such ``abstract'' \emph{representation}.)  Since the reflecting stations deserved ``places'' of their own, it was natural to represent them by vectors $s_1,s_2,s_3$, say, forming a basis of $V$.  Again, any basis would do.  Of course, $S$ itself would be assigned the zero vector.  Now, the reflecting stations could not be allowed to be totally arbitrary; they had to remain at fixed ``distances'' from $S$ and from one another.  But what are distances? These had to be physically determinable.  $S$ has only a clock- no measuring rods.  As in radar, one could define ``distance'' in terms of \emph{time interval}  through a parameter called the velocity of light, $c$.  By doing some further signalling,  involving various reflections, $S$ could obtain a set of 6 time intervals that would correspond to  6 transition times, $SS_1,SS_2,SS_3,S_1S_2, S_2S_3,S_3S_1$.  These multiplied by $c$ would be taken as the lengths or ``norms'' of the 6 vectors $s_1,s_2,s_3,s_1-s_2,s_2-s_3,s_3-s_1$.  These 6 numbers would then uniquely determine the scalar or inner product on $V$, thus making it into an inner product space.  Note that although $V$ and a basis for it were arbitrarily chosen, the scalar product  was determined by the observation system itself.  The problem of obtaining from the 5-tuple of an event a representing vector $p$, say, in $V$ was then a problem of linear algebra (see\cite{Agashe1} for details, with a slightly different notation).  There was a ``technical'' hitch, however.  Not any set of 6 numbers would do; this was explored in an ``addendum''\cite{Agashe2}.

The stage was set to admit another observation system,  an observer $S'$, say, with a clock and reflecting stations $S_1',S_2',S_3'$. This observer could choose a vector space, $V'$, say, not necessarily the same as $V$ of $S$, basis vectors $s_1',s_2',s_3'$ to represent its stations, and using the same ``velocity of light'' constant $c$, determine a scalar product on it, and finally obtain the representing vector $p'$, say, and time $t'$, say, of the same event for which $S$ had obtained $p$  as position vector and $t$ as time.  To relate $p,t$ with $p',t'$, one assumed, as usual, that the system $S'$ was in uniform motion relative to system $S$  with velocity $v$. This motion would be observed by $S$, and thus, $v$ would be a vector in $V$. $S$ would observe the motions of $S',S_1',S_2',S_3'$ to be given by the vectors $d_0+tv,d_0+tv+d_1,d_0+tv+d_2,d_0+tv+d_3$, say, $d_0,d_1,d_2,d_3$ being all vectors in $V$.  To go further, one needed some relation between the clocks of $S$ and $S'$ in the following sense.  Suppose that as $S'$ moves, the clocks of $S$ and $S'$ \emph{at} $S'$ show values $t$ and $t'$, respectively.  We need some relation between these two ``times''.  We assume, with Einstein, linearity of this relation: $t' = \beta_1t$, where $\beta_1$ is some constant.  This is the only assumption of linearity that we make.  We then prove(see\cite{Agashe1}) that the following linear relations hold between the times and places of the events in the two systems:
\begin{equation}
t' = \beta_1\left[t-\frac{(p-d_0-tv,v)_S}{c^2 - v^2}\right]
\end{equation}
where the symbol $(u,w)_S$ denotes the scalar product of the vectors $u$ and $w$ in $V$, and
\begin{equation}
p' = T (p-(d_0+tv))
\end{equation}
where $T$ is a linear transformation on $V$ onto $V'$ such that it maps each vector $d_i$ of $V$ to the vector $s_i'$ of $V'$, i.e., the vectors in $V$ representing the relative positions of the stations of $S'$ are mapped to the vectors in $V'$ representing the stations of $S'$.  This completes a summary of our derivation of the Lorentz transformation in\cite{Agashe1}. We call it ``Einstein's Lorentz transformation'' because we have followed an Einsteinian approach - except with respect to the meaning of ``position''.

\section{  LORENTZ TRANSFORMATION IN GALILEAN FORM}

Although we allowed the possibility that the representation vector spaces $V$ and $V'$ could be different, they could be chosen to be the same.  Further, the vectors $d_i$ representing the relative positions in $V$ of the stations of $S'$  could be chosen to be the basis vectors for $S'$. Thus, we could choose $s_i' = d_i$.  Of course, the scalar products could be different, as they are dictated by the observational data.  The transformation $T$ then becomes the identity transformation, and for the vectors representing the place of the event in $S$ and $S'$, we obtain the following simple \emph{Galilean} relation:
\begin{equation}
p' = p-(d_0 +tv)
\end{equation}
which can be seen as the Galilean position vector of $P$ relative to $S'$.  It must be emphasized, however, that there are still two representations because there are possibly two different scalar products for $S$ and $S'$, and these scalar products relate to two different calculations of distances in the two systems.  Both are \emph{Euclidean} in the sense they are both based on a scalar product.  The normal ``Euclidean'' co-ordinate systems use the distance $\sqrt{x^2 + y^2 + z^2}$, which is related to a special scalar product.

Let us consider in this context the Einstein form of the Lorentz equations: 
\begin{equation}
\label{NLT}
x' = \beta (x - vt), \qquad
y' = y, \qquad
z' = z, \qquad
t' = \beta (t-\frac{vx}{c^2}).
\end{equation}
One could argue that one could put them in the Galilean form by using new variables $\bar{x}, \bar{y}, \bar{z}$:%
\begin{equation}
\bar{x} = x -vt, \qquad
\bar{y}=y, \qquad
\bar{z} =z 
\end{equation}
and explicitly defining a new distance for $S'$ given in terms of norm by
\begin{equation}
 {||(p,q,r)|| _{S'}}^2 = \beta^2 p^2 + q^2  + r^2 ,
 \end{equation}
choosing the constant $\beta_1$ equal to $1/ \beta$, but this would have looked  like a mathematical ``trick''.  In contrast, here we have envisaged the possibility of a different scalar product and distance in the very notion of representation by a vector.

\section{REMARKS}

The Galilean form for the position vector has the advantage that the Einsteinian factor $\beta$ does not appear in it, making manipulations easier. Also, length contraction and time dilatation disappear.  However, $\beta$ does remain in a slightly different appearance in the equation relating the times.  The additional factor $\beta_1$ could be chosen as unity. The main point of this paper is, however, that the concept of ``position'' has to be defined and that there is a choice in representing position. The present paper could be read as a postscript to the earlier papers \cite{Agashe1} and \cite{Agashe2}. 

It would be interesting to apply the coordinate-free vector representation of position to Maxwell's equations and to see how the Galilean form of the Lorentz transformation works out.

\end{document}